\newif\iffigs\figsfalse

\catcode`\@=11
\def\slash{\mathpalette\make@slash}
\def\make@slash#1#2{\setbox\z@\hbox{$#1#2$}%
  \hbox to 0pt{\hss$#1/$\hss\kern-\wd0}\box0}
\catcode`\@=12 
\def\al{\alpha}
\def\pa{\partial}
\def\lap{\triangle}
\def\ov{\over}
\def\ha{{1\over 2}}
\def\>{\rangle}
\def\<{\langle}
\def\es{\!=\!}

\def\mtx#1{\quad\hbox{{#1}}\quad}
\iffigs
 \input epsf 
 \def\PSGraphic#1#2{%
 \def\epsfsize##1##2{#2##1}
 \vskip -30pt
 \centerline{\epsfbox{#1}}
 \vskip 10pt}
\else

\fi
\documentstyle[12pt]{article}
\begin{document}
\titlepage
\begin{flushright} ETH-93-38\\September  1993
\end{flushright}
\begin{center}\Large{{ Temperature- and Curvature
Dependence of the Chiral Symmetry
Breaking in 2D Gauge Theories}}
\end{center}

\vspace{1cm}
\begin{center}I. Sachs\footnote[1]{E-mail: ivo@itp.ethz.ch} and A.
Wipf\footnote[2]{E-mail: wipf@itp.ethz.ch}\\  Institute for
Theoretical Physics\\ Eidgen\"ossische Technische
Hochschule, H\"onggerberg\\ CH-8093 Z\"urich,
Switzerland\\
\end{center}

\vspace{3cm}

\begin{abstract} The partition function and the order parameter for
the chiral symmetry breaking are computed for a family of
2-dimensional interacting theories containing the gauged Thirring
model. In particular we derive non-perturbative expressions for
the dependence of the chiral condensate on the temperature
and the curvature.  Both, high temperature and high curvature supress
the condensate exponentially and we can associate an effective
temperature to the curvature.\end{abstract}
\newpage
\section{Introduction}
Despite of the considerable amount of work devoted to the subject of chiral
symmetry breaking in gauge theories and in particular $QCD$, the
understanding of this non-perturbative phenomenon is still
unsatisfactory \cite{1}. Also the behaviour of quantum systems in a
hot and dense enviroment (eg. in neutron
stars or in the early universe) are still under active
investigation \cite{1}.
On another front there has been much effort on the apparently
different problem of quantizing
self-interacting theories in a background gravitational field \cite{2}.\par
Rather than seeking new partial results for realistic $4$-dimensional
theories we analyse a family of interacting theories of charged
fermions, scalars,
pseudo-scalars and photons propagating in 2-dimensional curved spacetime
in detail. These models are defined by the action
\begin{eqnarray}
&S=\int\sqrt{-g}&\Big[\bar\psi i\gamma^\mu(D_\mu- ig_1\partial_\mu\lambda+ig_2
\eta_\mu^{\;\;\nu}\partial_\nu\phi )\psi\nonumber\\& &
+g^{\mu\nu}(\partial_\mu\phi\partial_\nu\phi+\partial_\mu\lambda\pa_\nu\lambda
)-g_3
{\cal R}\lambda-{1\over4}
F_{\mu\nu}F^{\mu\nu}\Big],\label{1} 
\end{eqnarray}
where $F_{\mu\nu}$ is the electromagnetic field strength and
$D_\mu\es\nabla_\mu-\!ieA_\mu$ the generally- and gauge covariant
derivative. This family
contains in particular the Schwinger model ($g_i\es0,\,i\es1,..,
3$)\cite{3} and the
gauged Thirring model ($g_1^2\es-g_2^2\es g^2,\,g_3\es0$)\cite{4,5} in
curved spacetime
$$
S_{Th}=\int\sqrt{-g}\Big[\bar\psi i\gamma^\mu D_\mu\psi
-\frac{g^2}{4}j^\mu j_\mu-{1\over4}F_{\mu\nu}F^{\mu\nu}\Big].
$$
The
coupling constant $g_3$ has been introduced in order to test the
effect of non-minimal coupling to the gravitational field. Finite
temperature effects are then included by quantizing the system on an
euclidean torus\footnote[1]{choosing a torus rather than a cylinder
provides us with an infrared regularization \cite{6}.} 
$[0,\beta]\times[0,L]$ with arbitrary metric. We choose coordinates such that
$$
g_{\mu\nu}=e^{2\sigma(x)}\pmatrix{|\tau|&0\cr0&1\cr},\mtx{where}
\tau=i\frac{\beta}{L}.
$$
$\beta$ is the inverse temperature and $L$ is the infrared cut-off
which will be removed after the correlation have been calculated.
Furthermore, finite
temperature boundary conditions are imposed on the quantum fields \cite{6}.
\par
On the torus a general gauge
potential with non-vanishing flux can be decomposed as
$$
A_\mu=A^k_\mu+t_\mu+\pa_\mu\al-\eta_{\mu\nu}\pa^\nu\varphi,
$$
where the last $3$ terms are recognized as Hodge decomposition
of the single valued part of $A$ and $A^k$ is an instanton potential
giving rise to a quantized flux
$
e\int F=2\pi\,k.
$
As a consequence the corresponding Dirac operator has $|k|$ zero
modes \cite{6} of chirality $sign\{k\}$. These zero modes are responsible for
a non vanishing chiral condensate $\langle\bar\psi\psi\rangle$ as can
be seen by inspecting the fermionic  generating functional \cite{6} in
the external
fields\footnote[2]{the harmonic field $h$ is needed for a consistent
quantization on the torus, analogous to the
harmonic part $t_\mu$ of the gauge field. On the torus the action
(\ref{1}) is changed to $S\rightarrow S+\int\sqrt{g}[g_2h_\mu j^\mu
+h_\mu h^\mu]$}$A,\phi,\lambda,h$ and sources $\eta,\bar\eta$
\begin{equation}
Z_F[A^k,\phi,\lambda,h_\mu,\eta,\bar\eta]=\prod\limits_{p=1}^{|k|}
(\bar\eta,\psi_{0p})(\psi_{0p}^{\dagger},\eta){\det}^\prime(i\slash
D)\,e^{-\int\sqrt{g}\bar\eta(x)S_e(x,y)\eta(y)}.\label{pf}
\end{equation}
Here $\psi_{0p}(x),\ p\es 1,...,|k|$, are the $|k|$ zero modes in the
topological sector $k$ and ${\det}^\prime(i\slash
D)$ denotes the zero mode truncated determinant. $S_e(x,y)$ is the
excited fermionic Green's function. We shall restrict ourselves to the
sectors $k\es0$  and $k\es1$, since these contribute to the partition
function and the chiral condensate, respectively. 
Finally we introduce a chemical potential for the conserved electric charge. 
In the euclidean forumlation this is done by shifting the zero
component of the gauge potential by an imaginary constant \cite{Actor}.\par
\section{Partition function}
As a first step in analyzing structure of the quantum theory we
evaluate the partition function formally defined by
\begin{equation}
Z_0=\int{\cal{D}}\big(A,\phi,\lambda,h\big)\,Z_F[A,\phi,\lambda,h]
e^{-S_B(A,\phi,\lambda,h)},\label{part}
\end{equation}
where
$$
{S_B(A,\phi,\lambda,h)}=-\int\sqrt{g}\phi\lap\phi+
\lambda\lap\lambda-h_\mu h^\mu-{1\over4}
F_{\mu\nu}F^{\mu\nu}.
$$
After a covariant gauge fixing (\ref{part}) is promoted to a well
defined quantity.
From (\ref{pf}) it is clear that only the trivial topological sector
contributes to the partition function. Then $Z_F[A,\phi,\lambda,h]$ equals
the determinant of the Dirac operator $\slash D$ which is related by
conformal- and chiral transformations to
$$
i\hat{\slash D}\equiv \hat\gamma^\mu(\pa_\mu-\frac{2\pi
i}{L}a_\mu),\;\mtx{where}\;
\frac{2\pi}{L}a_\mu=et_\mu+g_2h_\mu-\tau\mu\delta_{0,\mu}.
$$
Hatted quantities refer to flat metric and constant gauge potentials.
The chemical potential is contained in the last term in $a_\mu$.
Integrating the chiral- and conformal anomalies \cite{BVW} we find
\begin{equation}
\det(i\slash D)=\det(i\hat{\slash
D})\,\exp\big[{\frac{1}{24\pi}S_L+\frac{1}{2\pi}\int\sqrt{g}G
\frac{1}{\lap}G}\big],\label{anomaly}
\end{equation}
where
$$
S_L={\frac{1}{4}\int {\cal{R}}\frac{1}{\lap}{\cal{R}}}
$$
is the Liouville action and $G=g_2\varphi+e\phi$. One must be careful
in computing the hatted determinant since the gauge potential is
complex. This has been done in \cite{9}
with the result
\begin{equation}
\det(i\hat{\slash
D})=\frac{1}{\big|\eta(\tau)\big|^2}\Theta\Big[{a_1\atop
-a_0}\Big](0,\tau)\bar\Theta\Big[{\bar a_1\atop
-\bar a_0}\Big](0,\tau).\label{hd}
\end{equation}
The remaining functional integrals in (\ref{part}) turn out to be of
iterative Gaussian type and yield after substitution of (\ref{hd})
\begin{equation}
Z_0=\sqrt{2\pi V}\;{e\ov m_\gamma}{L\ov \beta\vert\eta(\tau)
\vert^4}
\;{1\over {\det}^{\prime\ha} (-\lap+m_\gamma^2)}
\,\exp\Big(({1\ov 12\pi}+g_3^2)\,S_L\Big)\label{part2}
\end{equation}
where $V\es\int\sqrt{g}$ and
$$
m_\gamma^2=\frac{e^2}{\pi}\frac{2\pi}{2\pi+g_2^2}
$$
is the dynamically generated "photon" mass.
This result already indicates that in the trivial topological sector
the theory (\ref{1}) should be equivalent to a free, massive, neutral,
boson even in curved space-time. Note that the mass depends on $g_2$.
In particular (\ref{part2}) shows that only the transversal part of the
current-current interaction contributes to the mass renormalization in
the Thirring model. Note also that the chemical potential does not
appear in the final result for the partition function. This may not
come as a surprise, because of the equivalence to a uncharged boson.
Also $\pa_\mu
Z[\mu]\es 0$ is the only result consistent with Gauss's law. We consider
this consistency as a confirmation of our definition of the fermionic
determinant which differs from previous ones in the literature
\cite{10}. The non-minimal coupling to gravity (for $g_3\neq 0$)
contributes to the gravitational anomaly and therefore affects the
intensity of the
Hawking radiation\par
\section{Chiral Condensate}
The chiral condensate $\<\bar\psi\psi\>$  is the order parameter for
the chiral symmetry breaking, responsible for the mass term in
(\ref{part2}). Here we evaluate the dependence of the order parameter
on temperature and curvature. Recalling
(\ref{pf}) we see that only configurations within the topological
sectors $k\es\pm 1$ can contribute to this expectation value. More precicely
\begin{equation}
\<\bar\psi(x) P_+\psi(x)\>=\frac{1}{Z_0}\int{\cal{D}}(\ldots)
\psi_{01}^{\dagger}(x)\psi_{01}(x){\det}^\prime(i\slash
D)\,e^{-S_B[A^1,\phi,\lambda,h]}\big|_{k=1},
\end{equation}
where $P_+=\ha(1+\gamma_5)$ is the projector on states with positive chirality.
$Z_0$ has been computed in the previous section (\ref{part2}). The
generalization
of (\ref{anomaly}) to non-zero $k$ reads
\begin{eqnarray}
{\det}^\prime(i\slash
D)&=&\det{{\cal N}_\psi
\ov \hat{\cal N}_\psi}\;{\det}^\prime(i\hat{\slash
D})
\exp\big(\frac{1}{24\pi}S_L\big)\nonumber\\& &
\cdot
\exp\big(\frac{1}{2\pi}\int\sqrt{g}G
\frac{1}{\lap}G+\frac{2k}{V}\int\sqrt{g}G+\frac{2\pi k^2}{\hat
V}\int\sqrt{\hat g}\chi\big),
\end{eqnarray}
where the hatted determinant now also contains the instanton potential.
${\cal{N}}$ is the normmatrix of the zero modes
$$
\psi^p_{0+}(x)=
e^{iF-\gamma_{5} (G+2k\pi\chi)-\ha\sigma}\hat\psi^p_{0+}(x)
$$
and $\chi(x)$
satisfies the differential equation
$$
\sqrt{g}\lap\chi=\sqrt{g}\frac{2\pi}{V}-\sqrt{\hat g}\frac{2\pi}{\hat V}.
$$
All information about the harmonics and the chemical potential is
contained in the zero modes. However,
$$
\int\,d^2t\;{\det}^\prime(i\hat{\slash
D})\psi_{01}^{\dagger}\psi_{01}=\frac{1}{\sqrt{2\beta L}}
$$
and hence
\begin{equation}
\langle \psi^\dagger P_+\psi\rangle=
\sqrt{-i\tau\ov \hat V}\,\vert \eta (\tau)\vert^2
e^{-2\pi^2/e^2V+2\pi/\hat V\int\sqrt{\hat g}\chi}
\Big\langle e^{-2(g\phi+e\varphi)(x)-\sigma(x)}
\Big\rangle_{ \phi\varphi},\label{cc}
\end{equation}
where the expectation value is evaluated with
$$
S_{eff}=\int\sqrt{g}\Big[\ha\varphi (\lap^2-{e^2\over\pi}\lap)\varphi
-{e^2\over\pi m_\gamma^2}\phi\lap\phi-{eg_2\over\pi}\phi\lap\varphi
\Big].
$$
A formal calculation of the resulting Gaussian integrals yields
\begin{eqnarray}
\langle \psi^\dagger P_+\psi\rangle&=&
\sqrt{-i\tau\ov \hat V}\vert \eta (\tau)\vert^2 e^{-2\pi^2/e^2V
+2\pi/\hat V\int\sqrt{\hat g}\chi}\,
e^{-\sigma(x)-4\pi\chi(x)}\nonumber\\
& &\cdot\exp\big[{2\pi^2m_\gamma^4\ov e^2}\,K(x,x)\big]
\exp\big[{2\pi g_2^2\over 2\pi+g_2^2}\,G_0(x,x)\big],\label{cc}
\end{eqnarray}
where
\begin{equation}
K(x,y)=\langle x\vert{1\over \lap^2-m_\gamma^2\lap}\vert y\rangle
={1\ov m_\gamma^2}\big(G_0(x,y)-G_{m_\gamma}(x,y)\big) \label{Kxx}
\end{equation}
and $G_{m},G_0$ are the massive and massless Green's functions
respectively.\par
As it stands (\ref{cc}) is still a formal expression since
$G_0(x,y)$ is logarithmically divergent when $x$ tends to $y$.
To extract a finite answer we need to renormalize the operator
$\exp(\al \phi)$. This wave function renormalization is equivalent
to the renormalization of the fermion field in the Thirring model
and thus is very much expected already in flat space time \cite{11}. Its
generalization to curved space-time is found to be
$$
G_0^{reg}(x,x)= -{1\over 2\pi}\log
\big[\frac{2\pi|\eta(\tau)|^2}{Lm_\gamma}\big].
$$
To determine the chiral condensate we also need to determine
$K(x,y)$ on the diagonal. In a first step we shall obtain
it for the flat torus. Its curvature dependence is then determined
in a second step.\par For $\sigma\es 0$ 
the Green's function $K$ has been computed in \cite{6}. Substitution
of this Green's function leads, after removing the infrared cut-off,
to the following exact formula for the {\it chiral condensate on flat space}
\begin{equation}
\langle\psi^\dagger P_+\psi\rangle_{\beta}=-
T\Big({m_\gamma\ov 2\pi T}\Big)^{g_2^2\ov 2\pi+g_2^2}
\exp\Big[-{\pi^2m_\gamma\ov e^2}T+{2\pi \ov 2\pi+g_2^2}F\Big],\label{chirt}
\end{equation}
where
$$
F(\beta)=\sum_{n>0}\Big[{1\over n}-{1\ov
\sqrt{n^2+(\beta m_\gamma/2\pi)^2}}\Big].
$$
For arbitrary values of the temperature and $g_2$ the infinite sum $F$
is evaluated on a computer (Fig.1). It is however interesting to
discuss some limiting cases.\par
For low temperatures, compared to $m_\gamma$ we have 
\begin{equation}
F(\beta)\to \gamma+\log\frac{\beta m_\gamma}{4\pi}+\frac{\pi}{\beta
m_\gamma},\label{Ft}
\end{equation}
where $\gamma\es 0.57721\dots$ is the Euler constant. Substitution of
(\ref{Ft})
yields the {\it zero temperature} result
\begin{equation}
\langle\psi^\dagger P_+\psi\rangle=
-{m_\gamma\ov 4\pi}\,2^{g_2^2/(2\pi+g_2^2)}\,
\exp\Big({2\pi\ov 2\pi+g_2^2}\gamma\Big)\hspace{.5cm}\mtx{for}
T\to 0.\label{smallT}
\end{equation}
On the other hand for temperatures large compared to the induced
photon mass $F$
vanishes. Thus we obtain the {\it high temperature behaviour}
\begin{equation}
\langle\psi^\dagger P_+\psi\rangle_{T}=
-T\Big({m_\gamma\ov 2\pi T}\Big)^{g_2^2\ov 2\pi+g_2^2}\,
\exp\Big(-{\pi^2 m_\gamma\ov e^2}T\Big)\mtx{for} T\to\infty.\label{bigT}
\end{equation}
Hence the chiral condensate decays exponentially for high temperatures
approaching zero assymptotically. The coupling to the pseudoscalars
$\phi$ weakens the effect of the temperature while the scalar field
$\lambda$ has no effect. For the gauged Thirring model this result
implies that only the transversal part of the current-current coupling
affects the chiral condensate.
Finally note that, as the partition function, the chiral condensate does
not depend on the chemical potential.\par
\iffigs
 \PSGraphic{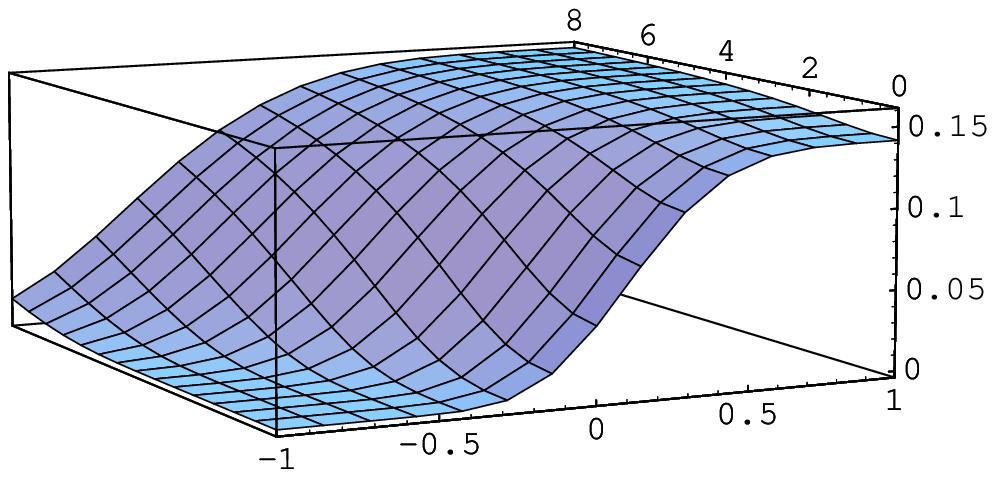}{1.0}
 \vspace{-2.5cm}
 \begin{center}{Fig. 1: The chiral condensate as a function of the temperature
and the coupling constant $g_2$.}
 \end{center}
 \vspace{-7.8cm}
$$\hspace{3cm}g_2$$
 \vspace{.5cm}
$$\hspace{12cm}\frac{\langle\psi^{\dagger}
P_+\psi\rangle}{m_\gamma}$$
 \vspace{2.1cm}
$$\hspace{2cm} \log_{10}\big[\frac{m_\gamma}{T}\big]$$
 \vspace{2cm}
\else
\message{No figures will be included. See TeX file for more
information.}
\fi
\par
How does the gravitational field affect
the chiral condensate? To answer this question we need to know the
massive Green's function, entering in (\ref{Kxx}), for non-trivial
gravitational fields (for simplicity we assume $T\es 0$). Let us first
consider  a space with constant positive curvature. Then $G_{m_\gamma}$ has
been computed explicitely \cite{13}. Here we only need the short
distance expansion, given by
\begin{equation}
G_{m_\gamma}(x,y)=-\frac{1}{4\pi}\big\{2\gamma+
\log\big(\frac{s^2{\cal{R}}}{8}\big)+\psi(\ha+\alpha)
+\psi(\ha-\alpha)+O(s^2)\big\},\label{GmR}
\end{equation}
where $\alpha^2\es\frac{1}{4}-\frac{2m_\gamma^2}{{\cal{R}}}$ and $\psi(z)$
is the Digamma function. Substituting (\ref{GmR}) into (\ref{Kxx}) we
end up with the exact formula for the {\it chiral condensate for
constant curvature} 
\begin{equation}
\langle\psi^{\dagger} P_+\psi\rangle_{\cal{R}}=\langle\psi^{\dagger}
P_+\psi\rangle_{{\cal{R}}=0}\cdot
\exp\Big[\frac{\pi}{2e^2}m_\gamma^2\big\{
\log\big(\frac{{\cal{R}}}{2m_\gamma^2}\big)
+\psi(\ha+\alpha)+\psi(\ha-\alpha)\big\}\Big].\label{chirR}
\end{equation}
The assymptotic expansions for {\it large-and small
curvatures} are easily worked out inserting the corresponding
expansions for the Digamma function \cite{14}. We find
\begin{equation}
\langle\psi^{\dagger} P_+\psi\rangle_{\cal{R}}=\langle\psi^{\dagger}
P_+\psi\rangle_{{\cal{R}}=0}\cdot
\exp\Big[-\frac{\pi}{12e^2}{\cal{R}}\Big]\hspace{.5cm}\hbox{for}
\;\frac{{\cal{R}}}{m_\gamma}\rightarrow 0\label{smallR}
\end{equation}
and
\begin{equation}
\langle\psi^{\dagger} P_+\psi\rangle_{\cal{R}}=\langle\psi^{\dagger}
P_+\psi\rangle_{{\cal{R}}=0}\cdot
\big(\frac{{\cal{R}}}{2m_\gamma^2}\big)^{\frac{\pi}{2\pi+g_2^2}}
\exp\Big[-\frac{\pi}{4e^2}
{\cal{R}}- \frac{\pi m_\gamma^2}{4e^2}\gamma\Big]\hspace{.5cm}\hbox{for}
\;\frac{{\cal{R}}}{m_\gamma}\rightarrow\infty.\label{bigR}
\end{equation}
Hence the chiral
condensate decays exponentially for large curvature analogous to the
high temperature behaviour. However, the pseudo-scalars do not supress
the effect of the curvature in contrast to (\ref{bigT}). Comparing the
exponentials in
(\ref{bigR}) to (\ref{bigT}) we are lead to define the curvature induced
effective temperature as
\begin{equation}
T_{eff}={{\cal R}\ov {4\pi m_\gamma}}.
\end{equation}
In passing we note that if we compare the prefactors, rather than the
exponentials, we would write
\begin{equation}
T_{eff}={{\cal R}^\ha\ov {4\pi\sqrt{2}}}.
\end{equation}
The latteridentification actually coincides (up to factor of $2$) with the
Hawking temperature of free scalars in de Sitter space \cite{15}.
The correct identification involves the (dynamical) mass of
the gauge field and is therefore not universal. From this observation
we learn that the temperature associated with curvature depends
on the matter content. Note finally that the non-minimal coupling
($g_3$) has no
effect on the chiral condensate. In Fig. 2 we have
plotted the chiral condensate for arbitrary constant values of the
curvature.\par 
\iffigs
 \PSGraphic{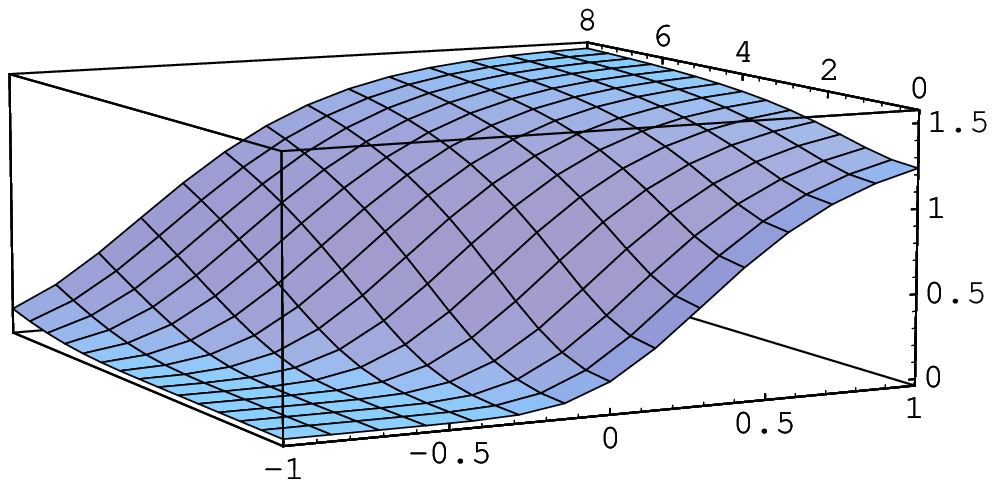}{1.0}
 \vspace{-2.5cm}
 \begin{center}{Fig. 2: The chiral condensate as a function of the curvature
and the coupling constant $g_2$.}
 \end{center}
 \vspace{-7.8	cm}
$$\hspace{3cm}g_2$$
 \vspace{.8cm}
$$\hspace{12cm}\frac{\langle\psi^{\dagger}
P_+\psi\rangle}{m_\gamma}$$
 \vspace{2.2cm}
 $$\hspace{1cm} \log_{10}\big[\frac{4\pi m_\gamma}{{\cal{R}}}\big]$$
 \vspace{1.5cm}
\else
 \message{No figures will be included. See TeX file for more
information.}
\fi
\par
For gravitational backgrounds with non-constant curvature we have to
refer to perturbative methods for the calculation of the massive
Green's function. Again we only need the short distance
expansion of $G_{m_\gamma}$. For geodesic distances $s$ small
compared to $m_\gamma^{-1}$ the massive
Green's function may be approximated by the Seeley DeWitt expansion
\cite{Christensen}
\begin{equation}
G_m(x,y)\sim {1\ov 4i}\sum_{j=0}^\infty a_j(x,y)
\big(-{\pa\ov \pa m^2}\big)^j\,H_0^{(2)}(ms),\label{H0}
\end{equation}
where $H_0^{(2)}$ is
the Hankel function of the second kind and order zero. In particular
$$
H_0^{(2)}(z)\to {2\ov i\pi}\big[\log{z\ov 2}+\gamma\big]
\mtx{for} z\to 0.
$$
Inserting (\ref{H0}) into (\ref{Kxx}) we end up with the following
expansion for the {\it chiral condensate in an arbitrary background}
\begin{equation}
\langle\psi^{\dagger} P_+\psi\rangle_{{\cal R}}=\langle\psi^\dagger
P_+\psi\rangle_{{{\cal R}}=0}\cdot
\exp\Big[-\frac{\pi}{2} \big({ m_\gamma\ov e}\big)^2 \sum_1^\infty
a_j(x){(j-1)!\ov m^{2j}}\Big],\label{chirS}
\end{equation}
where we have used that $a_0(x)\es1$. The first order
contribution involves $a_1(x)\es{1\over 6}{\cal R}$ and reproduces 
the assymptotic
behaviour (\ref{smallR}). Higher order contributions must be taken
into account to uncover the effect of variable curvature. For this one
has to substitute is the 
corresponding Seeley DeWitt coefficients $a_j$ into (\ref{chirS}). These
have been computed up to $j\es 5$ \cite{Seeley}.
\section{Summary}
We have computed the partition function and the order parameter of the
chiral symmetry breaking for a Thirring-like gauge theory. In particular
we find that both, high temperature and high curvature supress the
condensate exponentially. Comparing the two results, we defined a
curvature induced effective temperature which, unlike the Hawking
temperature, depends on the matter content and is therefore not
universal. Furthermore we have shown that a non-minimal
coupling to gravity affects the Hawking radiation while it has no
effect on the chiral symmetry breaking. The non-minimal coupling
choosen in our model is however not unique and this result is
therefore not general. Finally we obtain that a chemical potential for
the electric charge does affect neither the partition function, nor
the chiral condensate, in consitency with Gauss's law.
\section{Acknowledgements}
\par\noindent
This work has been supported by the Swiss National Science
Foundation. We would like to thank A.Dettki and H. M\"uller
for discussions.\par

\end{document}

\bye